
\input epsf
\newskip\tableskipamount \tableskipamount=10pt plus 4pt minus 4pt

\def\upar{\uparrow\kern-9.\exec1pt\lower.2pt \hbox{$\uparrow$}}

\catcode`\@=11
\def\unredoffs{\voffset=13mm \hoffset=8.5truemm} 
\def\redoffs{\voffset=-12.truemm\hoffset=-9truemm} 
\def\speclscape{}
\newbox\leftpage \newdimen\fullhsize \newdimen\hstitle \newdimen\hsbody
\newdimen\hdim
\hfuzz=1pt
\def\bigans{b }
\message{ big or little (b/l)? }\read-1 to\answ
\ifx\answ\bigans\message{(Format simple colonne 12pts.}
\magnification=1200 \unredoffs\hsize=141truemm\vsize=219truemm
\hsbody=\hsize \hstitle=\hsize 
\else\message{(Format double colonne, 10pts.} \let\l@r=L
\magnification=1000 \vsize=182.5truemm
\redoffs%
\hstitle=117.5truemm\hsbody=117.5truemm\fullhsize=258truemm\hsize=\hsbody 
\output={
  \almostshipout{\leftline{\vbox{\makeheadline\pagebody\makefootline}}}
\advancepageno%
}
\def\almostshipout#1{\if L\l@r \count1=1 \message{[\the\count0.\the\count1]}
      \global\setbox\leftpage=#1 \global\let\l@r=R
 \else \count1=2
  \shipout\vbox{\speclscape{\hsize\fullhsize}
      \hbox to\fullhsize{\box\leftpage\hfil#1}}  \global\let\l@r=L\fi}
\fi
\def\deqalignno#1{\displ@y\tabskip\centering \halign to
\displaywidth{\hfil$\displaystyle{##}$\tabskip0pt&$\displaystyle{{}##}$
\hfil\tabskip0pt &\quad
\hfil$\displaystyle{##}$\tabskip0pt&$\displaystyle{{}##}$ 
\hfil\tabskip\centering& \llap{$##$}\tabskip0pt \crcr #1 \crcr}}
\def\deqalign#1{\null\,\vcenter{\openup\jot\m@th\ialign{
\strut\hfil$\displaystyle{##}$&$\displaystyle{{}##}$\hfil
&&\quad\strut\hfil$\displaystyle{##}$&$\displaystyle{{}##}$
\hfil\crcr#1\crcr}}\,}
\newcount\nosection
\newcount\nosubsection
\newcount\neqno
\newcount\notenumber
\newcount\nofigure
\newif\ifappmode
\def\table{\jobname.tab}
\def\equation{\jobname.equ}
\def\equadefs{\jobname.eqd}
\newwrite\equa
\newwrite\tab 
\newwrite\eqdf
\newdimen\hulp
\def\maketitle#1{
\edef\oneliner##1{\centerline{##1}}
\edef\twoliner##1{\vbox{\parindent=0pt\leftskip=0pt plus 1fill\rightskip=0pt
plus 1fill 
                     \parfillskip=0pt\relax##1}} 
\setbox0=\vbox{#1}\hulp=0.5\hsize
                 \ifdim\wd0<\hulp\oneliner{#1}\else
                 \twoliner{#1}\fi}
\def\preprint#1{{\sacfont }\hfill{#1} \vskip 20mm}
\def\title#1\par{\gdef\titlename{#1}
\maketitle{\ssbx\uppercase\expandafter{\titlename}}
\vskip20truemm
\nosection=0
\neqno=0
\notenumber=0
\nofigure=0
\def\prefix{}
\appmodefalse
\mark{\the\nosection}
\message{#1}
\immediate\openout\equa=\equation
\immediate\openout\eqdf=\equadefs
}
\def\section#1\par{\vskip0pt plus.1\vsize\penalty-100\vskip0pt plus-.1
\vsize\bigskip\vskip\parskip
\message{ #1}
\ifnum\nosection=0\ifappmode\relax\else\immediate\openout\tab=\table%
\edef\ecrire{\write\tab{\par\noindent{\ssbx\ \titlename} 
\string\leaderfill{\noexpand\number\pageno}}}\ecrire\fi\fi
\advance\nosection by 1\nosubsection=0\neqno=0
\vbox{\noindent\bf{\prefix\the\nosection\ #1}}
\mark{\the\nosection}\bigskip\noindent
\xdef\ecrire{\write\tab{\string\par\string\item{\prefix\the\nosection}
#1
\string\leaderfill {\noexpand\number\pageno}}}\ecrire}
\def\appendix#1#2\par{\bigbreak\nosection=0
\appmodetrue
\notenumber=0
\neqno=0
\def\prefix{A}
\mark{\the\nosection}
\message{\appendixname}
\leftline{{\bf Appendices}} 
\leftline{\uppercase\expandafter{#1}}
\leftline{\uppercase\expandafter{#2}}
\bigskip\noindent\nonfrenchspacing
\edef\ecrire{\write\tab{\par\noindent{Appendices\ #1\
#2}\string\leaderfill{\noexpand\number\pageno}}}\ecrire}%
\def\subsection#1\par {\vskip0pt plus.05\vsize\penalty-100\vskip0pt
plus-.05\vsize\bigskip\vskip\parskip\advance\nosubsection by 1
\vbox{\noindent\it\prefix\the\nosection.\the\nosubsection\
#1}\smallskip\noindent 
\edef\ecrire{\write\tab{\string\par\string\itemitem
{\prefix\the\nosection.\the\nosubsection} {#1}
\string\leaderfill{\noexpand\number\pageno}}}\ecrire
} 
\def\note #1{\advance\notenumber by 1
\footnote{$^{\the\notenumber}$}{\sevenrm #1}} 

\parindent=1em 
\newinsert\margin
\dimen\margin=\maxdimen
\count\margin=0 \skip\margin=0pt
\def\nolabels{\def\wrlabel##1{}\def\eqlabel##1{}\def\reflabel##1{}}
\def\writelabels{\def\wrlabel##1{\leavevmode\vadjust{\rlap{\smash%
{\line{{\escapechar=` \hfill\rlap{\sevenrm\hskip.03in\string##1}}}}}}}%
\def\eqlabel##1{{\escapechar-1\rlap{\sevenrm\hskip.05in\string##1}}}%
\def\reflabel##1{\noexpand\llap{\noexpand\sevenrm\string\string\string##1}}}
\nolabels
\global\newcount\refno \global\refno=1
\newwrite\rfile
\def\ref{[\the\refno]\nref}
\def\nref#1{\xdef#1{[\the\refno]}\writedef{#1\leftbracket#1}%
\ifnum\refno=1\immediate\openout\rfile=\jobname.ref\fi
\global\advance\refno by1\chardef\wfile=\rfile\immediate
\write\rfile{\noexpand\item{#1\ }\reflabel{#1\hskip.31in}\pctsign}\findarg}
\def\findarg#1#{\begingroup\obeylines\newlinechar=`\^^M\pass@rg}
{\obeylines\gdef\pass@rg#1{\writ@line\relax #1^^M\hbox{}^^M}%
\gdef\writ@line#1^^M{\expandafter\toks0\expandafter{\striprel@x #1}%
\edef\next{\the\toks0}\ifx\next\em@rk\let\next=\endgroup\else\ifx\next\empty%
\else\immediate\write\wfile{\the\toks0}\fi\let\next=\writ@line\fi\next\relax}}
\def\striprel@x#1{}
\def\em@rk{\hbox{}} 
\def\semi{;\hfil\break}
\def\addref#1{\immediate\write\rfile{\noexpand\item{}#1}} 
\def\listrefs{\immediate\closeout\rfile\writestoppt\baselineskip=14pt%
\vskip0pt plus.1\vsize\penalty-100\vskip0pt plus-.1
\vsize\bigskip\vskip\parskip\centerline{{\bf References}}\bigskip%
{\frenchspacing%
\parindent=20pt\escapechar=` \input \jobname.ref\vfill\eject}%
\nonfrenchspacing}
\def\startrefs#1{\immediate\openout\rfile=\jobname.ref\refno=#1}
\def\xref{\expandafter\xr@f}\def\xr@f[#1]{#1}
\def\refs#1{[\r@fs #1{\hbox{}}]}
\def\r@fs#1{\ifx\und@fined#1\message{reflabel \string#1 is undefined.}%
\xdef#1{(?.?)}\fi \edef\next{#1}\ifx\next\em@rk\def\next{}%
\else\ifx\next#1\xref#1\else#1\fi\let\next=\r@fs\fi\next}
\newwrite\lfile
{\escapechar-1\xdef\pctsign{\string\%}\xdef\leftbracket{\string\{}
\xdef\rightbracket{\string\}}}

\def\writestop{\def\writestoppt{\immediate\write\lfile{\string\pageno%
\the\pageno\string\startrefs\leftbracket\the\refno\rightbracket%
\string\def\string\secsym\leftbracket\secsym\rightbracket%
\string\secno\the\secno\string\meqno\the\meqno}\immediate\closeout\lfile}}
\def\writestoppt{}\def\writedef#1{}
\def\eqnn{\global\advance\neqno by 1 \ifinner\relax\else%
\eqno\fi(\prefix\the\nosection.\the\neqno)}
\def\eqnd#1{\global\advance\neqno by 1 \ifinner\relax\else%
\eqno\fi(\prefix\the\nosection.\the\neqno)
{\xdef#1{($\prefix\the\nosection.\the\neqno$)}}
\edef\ewrite{\write\eqdf{\string\def\string#1{($\prefix\the\nosection.%
\the\neqno$)}}%
\write\eqdf{}}\ewrite%
\edef\ewrite{\write\equa{{\string#1},(\prefix\the\nosection.\the\neqno)
{\noexpand\number\pageno}}\write\equa{}}\ewrite}
\def\eqna#1{\global\advance\neqno by1
{\xdef #1##1{\hbox{$(\prefix\the\nosection.\the\neqno##1)$}}}
\edef\ewrite{\write\eqdf{\string\def\string#1{($\prefix\the\nosection.%
\the\neqno$)}}%
\write\eqdf{}}\ewrite%
\edef\ewrite{\write\equa{{\string#1},(\prefix\the\nosection.\the\neqno)
{\noexpand\number\pageno}}\write\equa{}}\ewrite}

\def\em@rk{\hbox{}} 
\def\xeqn{\expandafter\xe@n}\def\xe@n(#1){#1}
\def\xeqna#1{\expandafter\xe@na#1}\def\xe@na\hbox#1{\xe@nap #1}
\def\xe@nap$(#1)${\hbox{$#1$}}
\def\eqns#1{(\e@ns #1{\hbox{}})}
\def\e@ns#1{\ifx\und@fined#1\message{eqnlabel \string#1 n'est pas defini.}%
\xdef#1{(?.?)}\fi \edef\next{#1}\ifx\next\em@rk\def\next{}%
\else\ifx\next#1\xeqn#1\else\def\n@xt{#1}\ifx\n@xt\next#1\else\xeqna#1\fi
\fi\let\next=\e@ns\fi\next}
\font\sacfont=eufm10 scaled 1440
\font\ssbx=cmssbx10 
\font\tenbi=cmmib7 
\scriptfont4=\tenbi
\def\e{\mathop{\rm e}\nolimits}
\def\d{{\rm d}}
\def\frac#1#2{{\textstyle{#1\over#2}}}
\def\tr{{\rm tr}}
\def\to{\rightarrow}
\def\half{\frac{1}{2}}

\def\del{\partial}
\def\phib{{\vec\phi}}
\def\Vc{\Omega_c}
\newskip\tableskipamount \tableskipamount=10pt plus 4pt minus 4pt

\def\upar{\uparrow\kern-9.\exec1pt\lower.2pt \hbox{$\uparrow$}}

\rightline{TECHNION-PH-95-21}
\preprint{SPhT/95-132}

\vskip -1.5cm
\title{The O(N) vector model in the large N limit revisited:
multicritical points and double scaling limit}

\centerline{Galit~Eyal${}^{1a}$, Moshe~Moshe${}^{1b}$,
Shinsuke~Nishigaki${}^{1c}$ and Jean~Zinn-Justin${}^{2}$}
\medskip\medskip\medskip
{\baselineskip14pt\centerline{${}^1$ Department of Physics,
Technion - Israel Institute of Technology,}
\centerline{Haifa, ISRAEL}
\smallskip
\centerline{${}^2$ CEA-Saclay, Service de Physique Th\'eorique*}
\centerline{F-91191 Gif-sur-Yvette Cedex, FRANCE}
\footnote{}{1a~email: phr74ge@technion.technion.ac.il}
\footnote{}{1b~email: moshe@physics.technion.ac.il}
\footnote{}{1c~present address: The Niels Bohr Institute,
Blegdamsvej 17, DK-2100
\indent Copenhagen O $\!\!\!\!\!\!/\,$, Denmark;~~~
email:~shinsuke@alf.nbi.dk}
\footnote{}{2~email: zinn@amoco.saclay.cea.fr}
\footnote{}{${^*}$Laboratoire de la Direction des
Sciences de la Mati\`ere du
Commissariat \`a \indent \ l'Energie Atomique}
\vskip6mm

\nref\rDavidetal{F. David, {\it Nucl. Phys.} B257 [FS14] 
(1985) 45, 543; ~~~J. Ambj{\o}rn, 
B. Durhuus and J. Fr\"ohlich, {\it Nucl. Phys.} B257 [FS14]
(1985) 433; ~V.A. Kazakov, {\it Phys. Lett.} B150 (1985) 282.}
\nref\rDS{M.
Douglas and S. Shenker, {\it Nucl. Phys.} B335 (1990) 635; 
~E. Br\'ezin and V. Kazakov, 
{\it Phys. Lett.} B236 (1990) 144; ~D. Gross and A.A. Migdal,
{\it Phys. Rev. Lett.} 64 (1990) 127; Nucl. Phys. B340 (1990) 333.}
\nref\rD{M. R.
Douglas, {\it Phys. Lett.} B238 (1990) 176; ~F. David, 
{\it Mod. Phys. Lett.} A5
(1990) 1019; ~P. Ginsparg, M. Goulian, M.
R. Plesser, and J. Zinn-Justin, {\it Nucl. Phys.} B342 (1990) 539;  ~A.
Jevicki and T. Yoneya,  {\it Mod. Phys. Lett.} A5 (1990) 1615.}
\nref\roneD{P.
Ginsparg and J. Zinn-Justin, {\it Phys. Lett.} B240 (1990) 333;
 ~E. Br\'ezin, V. A. Kazakov, and Al. B. Zamolodchikov,
{\it Nucl. Phys.} B338 (1990) 673;
 ~G. Parisi, {\it Phys. Lett.} B238 (1990) 209, 213;
{\it Europhys. Lett.} 11 (1990) 595;
 ~D. J. Gross and N. Miljkovic, {\it Phys. Lett.} B238 (1990) 217;
 ~For reviews see P. Di Francesco, P. Ginsparg and J. Zinn-Justin,
{\it Phys. Rep.} 254 (1995) 1.}
\nref\rAMP{
S. Nishigaki and T. Yoneya, {\it Nucl. Phys.} B348 (1991) 787;
~A. Anderson, R.C.
Myers and V. Periwal, {\it Phys. Lett.} B254 (1991) 89, {\it Nucl. Phys.}
B360 (1991) 463;
~P. Di Vecchia, M. Kato and N. Ohta, {\it Nucl. Phys.} B357 (1991) 495.}
\nref\rVecd{J. Zinn-Justin, {\it Phys. Lett.} B257 (1991) 335\semi
P. Di Vecchia, M. Kato and N. Ohta, {\it Int. J. Mod. Phys.} A7
(1992) 1391; ~T. Yoneya, {\it Prog. Theor. Phys. Suppl.} 107 (1992) 229.}

\centerline{\bf ABSTRACT}
\vskip4mm
The multicritical points of the $O(N)$ invariant $N$ vector model in the
large $N$ limit are reexamined. Of particular interest are the subtleties
involved in the stability of the phase structure at critical dimensions.
In the limit $ N \to \infty$ while the coupling $g \to g_c$  
in a correlated manner (the double scaling limit) a massless
bound state $O(N)$ singlet is formed
and powers of $1/N$ are compensated by IR singularities. The 
persistence of the $N \to \infty$ results beyond the leading
order is then studied with particular interest in the
possible existence of a phase
with propagating small mass vector fields and
a massless singlet bound state. We point  out that 
under certain conditions the
double scaled theory  of the singlet field 
is non-interacting  in critical dimensions. 
\vskip4mm
\section{Introduction}

Statistical mechanical properties of random surfaces as well as randomly
branched polymers can be analyzed within the framework of large $N$ expansion.
In the same manner in which matrix models in their double scaling limit  
\refs{\rDavidetal,\rDS,\rD,\roneD} provide representations  of 
dynamically triangulated random
surfaces summed on different topologies,  
$O(N)$ symmetric vector models represent discretized branched polymers
in this limit, where
$N\to\infty$ and the coupling constant $g \to g_{c}$ in a correlated 
manner \refs{\rAMP,\rVecd}. The surfaces in the case of matrix models, and  
the randomly branched polymers 
in the case of vector models are classified by the 
different topologies of their
Feynman graphs and thus by powers of 1/N.  Though  matrix theories attract
most  attention, a detailed understanding of these theories exists
only for dimensions  $d \leq 1$. On the other hand,
in  many cases, the $O(N)$ vector models can be successfully studied  also  in
dimensions $d > 1$ \rVecd, and thus, provide us with intuition for the search 
for a possible description of quantum field theory
in terms of extended objects in four dimensions,
which is a long lasting problem in elementary 
particle theory.  

The double scaling limit in $O(N)$ vector quantum field theories 
reveals an interesting phase structure beyond $N\to \infty$ limit.
In particular, though the $ N  \to \infty $ multicritical  structure 
of these models is generally well understood, there are certain cases where
it is still unclear which of the features survives at finite $N$, and to what
extent. One such problem is the multicritical behavior
of $O(N)$ models {\it at critical dimensions}
\ref\rBMB{W.A. Bardeen, M. Moshe, M. Bander, 
{\it Phys. Rev. Lett.} 52 (1984) 1188.}.
Here, one finds that in the $N \to \infty$ limit, 
there exists a non-trivial UV fixed point, scale invariance
is spontaneously broken, and the one parameter family of ground
states contains a massive vector and a massless bound state, a Goldstone 
boson-dilaton.
However, since it is unclear whether this structure is likely to survive
for finite $N$
\ref\rDKN{F. David, D.A. Kessler and H. Neuberger, {\it Phys. Rev. Lett.} 53
(1984) 2071, {\it Nucl. Phys.} B257 [FS14] (1985) 695.}, 
one would like to know whether 
it is possible to construct a local field theory of a massless 
dilaton via the double scaling limit, 
where all orders in
$1/N$ contribute. The double scaling limit is viewed as the limit at which the
attraction between the $O(N)$ vector quanta reaches a value at $g \to g_c$,
at which a massless bound state
is formed in the $N \to \infty$ limit, while the 
mass of the vector particle stays finite. In this
limit, powers of $1/N$ are compensated by IR singularities and thus all
orders in $1/N$ contribute.
\par
In section 2 the double scaling limit for simple integrals 
and quantum mechanics is recalled, introducing a formalism which will be
useful for field theory examples. \par
In section 3 the special case of field theory in dimension two is
discussed, slightly generalizing previous established results.\par
In higher dimensions a new phenomenon arises: the possibility of a spontaneous
breaking of the $O(N)$ symmetry of the model, associated to the Goldstone
phenomenon. 

Before discussing a possible double scaling limit,
the critical and multicritical points of the $O(N)$ vector model are
reexamined in section 4. In particular, a certain sign ambiguity that
appears in the expansion of the gap equation is noted, and related to the
existence of the IR fixed point in dimensions $2<d<4$. In section 5 
we discuss the subtleties and conditions for the existence of an $O(N)$ singlet
massless bound state 
along with a small mass  $O(N)$ vector particle excitation. It is pointed out
that the correct massless effective field theory is obtained after the
massive $O(N)$ scalar is integrated out. Section 6 is devoted to 
the double scaling limit with a particular emphasis on this limit in theories
at their critical dimensions. In section 7 the main conclusions are
summarized. 

\section Double scaling limit: simple integrals and quantum mechanics

The double scaling limit 
\refs{\rDavidetal,\rDS,\rD,\roneD} of the vector model has already been
investigated \rVecd, for dimensions $d\le1$ and for the simple $(\phib^2)^2$
field theory.  
We first recall results obtained in $d=0$ and $d=1$ dimensions, dimensions in
which the matrix models have equally been solved. We however introduce a
general method, not required here, but useful in the general field theory
examples. 
\subsection The zero dimensional example

Let us first recall the zero dimensional example. The partition function $Z$
is given by
$$\e^Z=\int\d^N\phib\exp\left[-NV\left(\phib^2\right)\right] .$$
The simplest method for discussing the large $N$ limit is of course to
integrate over angular variables (see appendix A1). Instead we introduce two
new variables $\lambda,\rho$ and use the identity
$$\exp\left[-N V(\phib^{2})\right] \propto \int
d\rho\,\d\lambda\exp\left\{-N\left[\half\lambda\left(\phib^{2}-\rho\right) +
V(\rho) \right]\right\}.\eqnd\egenideno $$
The integral over $\lambda$ is really a Fourier representation of
a $\delta$-function and thus the contour of integration runs parallel to the
imaginary axis. The identity
\egenideno\ transforms the action into a quadratic form in $\phib$.
Hence the integration over $\phib$ can be  performed and the
dependence in $N$ becomes explicit
$$\e^Z \propto 
 \int d\rho\,\d\lambda\exp\left\{-N\left[-\half\lambda\rho + 
V(\rho)+\half \ln\lambda \right]\right\}. $$
The large $N$ limit is obtained by steepest descent. The saddle point is given
by 
$$V'(\rho)=\half\lambda\,,\qquad \rho=1/\lambda\,.$$
The leading contribution to $Z$ is proportional to $N$ and obtained by
replacing $\lambda,\rho$ by the saddle point value. 
The leading correction is obtained by expanding $\lambda,\rho$ around the
saddle point and performing the gaussian integration. It involves the
determinant $D$ of the matrix $\bf M$ of second derivatives
$$ {\bf M}=\pmatrix{-\half\lambda^{-2} & -\half \cr -\half &
V''(\rho) \cr} \,,\quad D=\det{\bf
M}=-\half\left(V''(\rho)/\lambda^2+\half\right).$$  
In the generic situation the resulting contribution to $Z$ is $-\half \ln D$.
However if the determinant $D$ vanishes the leading order integral is no
longer gaussian, at least for the degree of freedom which corresponds to the
eigenvector with vanishing eigenvalue. The condition of vanishing of the
determinant also implies that two solutions of the saddle point equation
coincide and thus corresponds to a surface in the space of the coefficients of
the potential $V$ where the partition function is singular (see appendix A1
for details).  \par
To examine the corrections to the leading large $N$ behaviour it remains
however possible to integrate over one of the variables by steepest descent.
At leading order this corresponds to solving
the saddle point equation for one
of the variables, the other being fixed. Here it is convenient to eliminate
$\lambda$ by the equation $\lambda=1/\rho$. One finds
$$\e^Z
\propto
\int d\rho\,\exp\left[-N\bigl(V(\rho)-\half\ln\rho\bigr)+O(1)\right].
$$ 
In the leading term we obviously recover
the result of the angular integration 
with $\rho=\phib^2$. For $N$ large the leading contribution arises from the
leading term in the expansion of $W(\rho)=V(\rho)-\half\ln\rho$ near the
saddle point: 
$$W(\rho)-W(\rho_s)\sim\frac{1}{n!}W^{(n)}(\rho_s)(\rho-\rho_s)^n.$$
The integer $n$ characterizes the nature of the critical point.
Adding relevant perturbations $\delta_k V$ of parameters $v_k$ to
the critical potential 
$$\delta_k V= v_k (\rho-\rho_s)^k ,\quad  1\le k\le n-2\, $$
(the term $k=n-1$ can always be eliminated by a shift of $\rho$)
we find the partition function at leading order for $N$
large in the scaling region: 
$$\e^{Z(\{u_k\})}\propto\int\d z\,\exp\left(-z^n-\sum_{k=1}^{n-2}u_k
z^k\right),$$ 
where $z\propto N^{1/n}(\rho-\rho_s)$ and 
$$u_k \propto N^{1-k/n}v_k $$
is held fixed.
\subsection Quantum mechanics

The method we have used above immediately generalizes to quantum mechanics,
although a simpler method involves the Schr\"odinger equation. We consider the
euclidean action
$$S(\phib)=N\int\d
t\,\left[\half\bigl(\d_t\phib(t)\bigr)^2+V(\phib^2)\right]. \eqnn $$ 
Note the unusual field normalization, the factor $N$ in front of the
action simplifying all expressions in the large $N$ limit.\par
To explore the large $N$ limit one has to take the scalar function $\phib^2$,
which self-averages, as a dynamical variable \ref\rbook{see for example
J. Zinn-Justin, {\it Quantum Field Theory and Critical Phenomena}, Oxford
University Press (Oxford 1989), second edition (Oxford 1993) chap.\ 28.}.
At each time $t$ we thus perform the transformation \egenideno. One
introduces two paths $\rho(t),\lambda(t)$ and writes
$$\eqalignno{&\exp\left[-N\int \d t\,V(\phib^{2
})\right]&\cr&\hskip10truemm \propto \int \left[\d\rho(t)\,\d\lambda(t)
\right] \exp\left\{-N\int \d t\left[\half\lambda\left(\phib^{2}-\rho\right) +
V(\rho) \right]\right\}.\hskip12truemm&\eqnd\egeniden\cr} $$
The integral over the path $\phib(t)$ is then gaussian and can be performed.
One finds 
$$e^Z=\int\d\rho(t)\d\lambda(t)\,\exp\left[-S_{\rm
eff}(\lambda,\rho)\right]\,$$ 
with
$$S_{\rm eff}=N\int\d t\,\left[-\half \lambda\rho+V(\rho)+\half \tr\ln
\bigl(-\d_t^2+\lambda(t) \bigr)\right].$$
Again, in the large $N$ limit the path integral can be calculated by steepest
descent. The saddle points are constant paths solution of
$$V'(\rho)=\half\lambda\,,\qquad \rho={1\over2\pi}\int{\d\omega\over
\omega^2+\lambda}={1\over2\sqrt{\lambda}} \,,$$
where 
$\omega$ is the Fourier energy variable conjugate to $t$.
Again a critical point is defined by the property that at least two
solutions to the saddle point equations coalesce. This happens when 
the determinant of the matrix of second derivatives of the equations vanishes:
$$\det\pmatrix{V''(\rho) & -\half \cr -\half & \-{1\over4\pi}\int{\d\omega
\over(\omega^2+\lambda)^2} \cr} =0\,.$$ 
The leading correction to the saddle point contribution is given by a gaussian
integration. The result involves determinant of the operator second derivative
of $S_{\rm eff}$. By Fourier transforming time the operator becomes a tensor
product of $2\times2$ matrices with determinant $D(\omega)$
$$D(\omega)=\det\pmatrix{V''(\rho) & -\half \cr -\half &
\-{1\over4\pi}\int{\d\omega'
\over(\omega'{}^2+\lambda)[(\omega-\omega')^2+\lambda]} \cr} 
\,.$$ 
Thus, the criticality condition is equivalent to $D(0)=0$. 
When the criticality condition is satisfied, the leading correction is no
longer given by steepest descent. Again, since at most one mode can be
critical, 
we can integrate over one of the path by steepest descent, which means 
solving the saddle point equation for one function, the 
other being fixed.
While the $\rho$-equation remains local, the $\lambda$ is now non-local,
involving the diagonal matrix element of the inverse of the differential
operator $-\d^2_t+\lambda(t)$. We shall see in next section how this problem
can be overcome in general. A special feature of quantum mechanics, however,
is that the determinant can be calculated, after a simple change of variable.
We set
$$\lambda(t)=\dot s(t)+s^2(t), $$
in such a way that the second order differential operator factorizes
$$-\d^2_t+\lambda(t)=-\bigl(\d_t+s(t)\bigr)\bigl(\d_t-s(t)\bigr).\eqnd\echgvar
$$
The determinant of a first order differential operator can be calculated
$$\ln\det\bigl(-\d^2_t+\lambda(t)\bigr)=\int\d t\, s(t) .$$
The jacobian of the transformation \echgvar~contributes at higher order in
$1/N$ and can be neglected. Therefore the effective action becomes
$$\eqalign{S_{\rm eff}&=N\int\d t\,\left[-\half (\dot s+s^2)\rho+V(\rho)+\half
s(t) \right] \cr
&=N\int\d t\,\left[-\half \rho s^2 +\half s(\dot\rho+1)+V(\rho)\right].\cr}$$ 
We can now replace $s$ by the solution of a local saddle point equation:
$$-s\rho+\half (\dot\rho+1)=0\,,$$
and find
$$S_{\rm eff}=N\int\d t\,\left[{\dot\rho^2\over8\rho}+{1\over8\rho}+
V(\rho)\right].$$
We recognize the action for the large $N$ potential at zero angular momentum
in the radial coordinate $\rho(t)=\phib^2(t)$. Critical points then are
characterized by the behaviour of the potential $W(\rho)$ 
$$W(\rho)=V(\rho)+{1\over8\rho}\,,$$
near the saddle point $\rho_s$
$$W(\rho)-W(\rho_s)\sim W^{(n)}(\rho_s){\left(\rho-\rho_s\right)^n\over
n!}\,.$$
At critical points the ground state energy, after subtraction of the classical
term which is linear in $N$, has a non-analytic contribution. To eliminate
$N$ from the action we set
$$t\mapsto t N^{(n-2)/(n+2)},\qquad \rho(t)-\rho_s\mapsto N^{-2/(n+2)}z(t). $$
We conclude that the leading correction to the energy levels is proportional
to $N^{-(n-2)/(n+2)}$. Note also that the scaling of time implies that higher
order time derivatives would be irrelevant, an observation which can be used
more  directly to expand the determinant in local terms, and will be important
in next section.\par
If we add relevant corrections to the potential
$$\delta_k V=v_k (\rho-\rho_s)^k ,\quad 1\le k\le n-2\,,$$
the coefficients $v_k$ must scale like
$$v_k\propto N^{2(k-n)/(n+2)}.$$
\section{The 2D $V(\phib \sp{2}) $ field theory in the double scaling limit}

In the first part we recall the results concerning the $O(N)$ symmetric
$V(\phib^2)$ field theory, where $\phib$ is $N$-component field, in the large
$N$ limit in dimension two because phase transitions occur in higher
dimensions, a problem which has to be considered separately. The action
is:
$$S(\phib)= N\int\d^2 x \left\{\half \left[ \partial_{\mu} \phib (x)
\right]^{2} +V\left(\phib^2\right) \right\} ,\eqnd{\eactONgii}$$
where an implicit cut-off $\Lambda$ is always assumed below. Whenever the
explicit dependence in the cut-off will be relevant we shall assume a
Pauli--Villars's type regularization, i.e.\  the replacement in action
\eactONgii\ of $-\phib\del^2\phib$ by
$$-\phib\del^2 D(-\del^2/\Lambda^2)\phib\,,\eqnd\eDPV$$
where $D(z)$ is a positive non-vanishing polynomial with $D(0)=1$.\par
As before one introduces two fields
$\rho(x)$ and $\lambda(x)$ and uses the identity \egeniden.
The effective action is then:
$$S_{\rm eff}=N\int\d^2 x \left[V(\rho)-\half \lambda
\rho \right]+\half N \tr\ln(-\Delta+\lambda ).\eqnd\eactefNgii $$
Again for $N$ large we evaluate the integral by steepest descent.
Since the saddle point value $\lambda$ is the
$\phib$-field mass squared, we set in general $\lambda=m^2$. With this
notation the two equations for the saddle point $m^2,\rho_s=\langle \phib^2
\rangle$ are: 
\eqna\esaddNgenii
$$\eqalignno{
V'(\rho_s)&=\half m^2\,,&\esaddNgenii{a}\cr
\rho_s&={1\over (2\pi)^2 }\int^{\Lambda} {\d^2 k \over k^2
+m^2}\,, & \esaddNgenii{b} \cr}$$
where we have used a short-cut notation
$${1\over (2\pi)^2 }\int^\Lambda {\d^2 k \over k^2
+m^2}\equiv {1\over (2\pi)^2 }\int{\d^2 k \over D(k^2/\Lambda^2) k^2
+m^2}\equiv B_1(m^2).$$
For $m\ll\Lambda$ one finds
$$\eqalign{B_1(m^2)&={1\over2\pi}\ln(\Lambda/m)+{1\over4\pi}\ln (8\pi C)
+O(m^2/\Lambda^2) \cr 
\ln (8\pi C) &= \int_0^\infty\d s\left( {1\over D(s)}-\theta(1-s)\right). \cr}
$$
As we have discussed in the case of quantum mechanics a critical point
is characterized by the vanishing at zero momentum of the  
determinant of second derivatives of the action at the saddle point. 
The mass-matrix has then a zero eigenvalue which, in field theory, corresponds
to the appearance of a new massless excitation other than $\phib$. In order to
obtain the effective action for this scalar massless mode 
we must integrate over one of the fields
 \ref\Ruhl{This point has also been pointed out by
J. Maeder and W. R\"uhl, hep-th/9505131, Kaiserslautern preprint.}. In the
field theory case the resulting effective action can no longer be written in
local form.  
To discuss the order of the critical point, however, we only need the action
for space independent fields, and thus for example we can eliminate $\lambda$
using the $\lambda$ saddle point equation. 

The effective $\rho$ potential $W(\rho)$ then reads
$$W(\rho)=V(\rho)-\half\int^{\lambda(\rho)}\d \lambda' \,\lambda'
{\del \over \del \lambda'} B_1(\lambda'),\eqnd\eWeffrho$$
where at leading order for $\Lambda$ large 
$$\lambda(\rho)=8\pi C \Lambda^2\e^{-4\pi\rho}. $$
The second term in Eq.~\eWeffrho ~in fact is the  kinetic energy
contribution to the  ground state free energy as can be viewed in
an Hartree--Fock variational calculation
that becomes exact  in the limit of $N \to \infty$
(see for example Ref.
 \ref\BarMo{W. A. Bardeen and M. Moshe
{\it Phys. Rev.  D} 28 (1983) 1372}   )  . The expression 
for the effective action in Eq.~ \eWeffrho~is
correct  for any $d$ and will be used  also in section 6.

Here we have:
$$W(\rho)=V(\rho)+C\Lambda^2\e^{-4\pi\rho}=V(\rho)+\frac{1}{8\pi}m^2
\e^{-4\pi(\rho-\rho_s)}.$$
A multicritical point is defined by the condition
$$W(\rho)-W(\rho_s)=O\left((\rho-\rho_s)^n\right)\eqnd\eWcrit.$$
This yields the conditions:
$$V^{(k)}(\rho_s)=\half (-4\pi)^{k-1} m^2\quad{\rm for}\ 1\le k\le n-1\,.$$
Note that the coefficients $V^{(k)}(\rho_s)$ are the coupling
constants renormalized at leading order for $N$ large.
If $V(\rho)$ is a polynomial of degree $n-1$ (the minimal polynomial model)
the  multicritical condition in Eq.~\eWcrit~
 and Eq.~\esaddNgenii{b}  
determines the critical values of  $n-1$
renormalized coupling constants as well as $\rho_s$
 in terms of $m^2$. 
\par 
When the fields are space-dependent it is simpler to eliminate $\rho$ instead,
because the corresponding field equation: 
$$V'\bigl(\rho(x)\bigr)=\half \lambda(x). \eqnd\esadroge $$
is local. This equation can be solved by expanding $\rho(x)-\rho_s$ in a power
series in $\lambda(x)-m^2$:
$$\rho(x)-\rho_s= {1\over2 V''(\rho_s)}\bigl(\lambda(x)-m^2\bigr)
+O\left((\lambda-m^2)^2\right) . \eqnd\esolrola $$
The resulting action for 
the field $\lambda(x)$ remains non-local but because, as we
shall see,  adding powers of $\lambda$ 
as well as adding derivatives make
terms less relevant, only the few first terms of a local expansion of the
effective action will be important.\par
If in the local expansion of the determinant we keep only the two
first terms we obtain an action containing at leading order a kinetic
term proportional to $(\del_\mu\lambda)^2$ and the interaction
$(\lambda(x)-m^2)^n$:
$$S_{\rm eff}(\lambda) \sim N\int\d^2 x\left[{1\over96\pi
m^4}(\del_\mu\lambda)^2 + \frac{1}{n!}S_n \bigl(\lambda(x)-m^2)^n\right],$$
where the neglected terms are of order $(\lambda-m^2)^{n+1}$, $\lambda\del^4
\lambda$, and $\lambda^2\del^2\lambda$ and 
$$S_n=W^{(n)}(\rho_s)[2V''(\rho_s)]^{-n}=W^{(n)}(\rho_s)(-4\pi m^2)^{-n} .$$
Moreover we note that together with the cut-off $\Lambda$, $m$ now also
acts as a cut-off in the local expansion.\par
To eliminate the $N$ dependence in the action we have, as in the example
of quantum mechanics, to rescale both the field $\lambda-m^2$ and coordinates:
$$\lambda(x)-m^2=\sqrt{48\pi}m^2 N^{-1/2}\varphi(x)\,,\quad x\mapsto
N^{(n-2)/4}x \,.
\eqnd\elamrescal $$
We find
$$S_{\rm eff}(\varphi)
\sim\int\d^2x\left[\half(\partial_\mu\varphi)^2+\frac{1}{n!} g_n
\varphi^n\right] .$$
In the minimal model, where the polynomial $V(\rho)$ has exactly degree $n-1$,
we find $g_n=6(48\pi)^{(n-2)/2}m^2$. 

As anticipated we observe that derivatives and powers of $\varphi$
are affected by negative powers of $N$, justifying a local expansion.
However we also note that the cut-offs ($\Lambda$  or the mass $m$) are now
also multiplied by $N^{(n-2)/4}$. Therefore the large $N$ limit also
becomes a large cut-off limit.
\medskip
{\it Double scaling limit.} The existence of a double scaling limit
relies on the existence of IR singularities due to the massless or small mass
bound state which can compensate the $1/N$ factors appearing in the large $N$
perturbation theory. \par 
We now add to the action relevant perturbations:
$$\delta_k V=v_k(\rho(x)-\rho_s)^k,  \quad 1\le k\le n-2. $$
Namely, adding
to the $\lambda$ action a sum of terms
proportional to $\int\d^2x(\lambda-m^2)^k$:
$$\delta_k S_{\rm eff}(\lambda)=N S_k \int\d^2x\,(\lambda-m^2)^k ,$$
where the coefficients $S_k$ are functions of the coefficients $v_k$.
After the rescaling \elamrescal~
these terms become
$$\delta_k S_{\rm eff}(\varphi)=\frac{1}{k!} g_k
N^{(n-k)/2}\int\d^2x\,\varphi^k(x)  \quad 1\le k\le n-2  $$
However, unlike quantum mechanics, it is not sufficient to scale the
coefficients $g_k$ with the power $N^{(k-n)/2}$ to obtain a finite scaling
limit. Indeed 
perturbation theory is affected by UV divergences, and we have just noticed
that the cut-off diverges with $N$. In two dimensions the nature of
divergences is very simple: it is entirely due to the self-contractions
of the interaction terms and only one divergent integral appears:
$$\left<\varphi^2(x)\right>={1\over4\pi^2}\int {\d^2 q\over q^2+\mu^2}\,,$$
where $\mu$ is the small mass of the bound state, required as an IR cut-off to
define perturbatively the double scaling limit.
We can then extract the $N$ dependence
$$\left<\varphi^2(x)\right>={1\over8\pi}(n-2)\ln N+O(1) .$$
Therefore the coefficients $S_k$ have also to cancel these UV divergences,
and therefore have a logarithmic dependence in $N$ superposed to the natural
power obtained from power counting arguments. In general for any potential
$U(\varphi)$
$$U(\varphi)=:U(\varphi):+\left[\sum_{k=1}{1\over2^k
k!}\left<\varphi^2\right>^k \left(\partial\over\partial \varphi\right)^{2k}
\right]:U(\varphi):\,,$$
where $:U(\varphi):$ is the potential from which self-contractions have been
subtracted (it has been normal-ordered). For example for $n=3$
$$\varphi^3(x)=:\varphi^3(x):+3\left<\varphi^2\right> \varphi(x) ,$$
and thus the double scaling limit is obtained with the behaviour
$$N g_1+{1\over16\pi}\ln N\quad {\rm and}\quad N^{1/2} g_2 \ {\rm fixed}\ .$$
For another example $n=4$ 
$$ g_1 N^{3/2}\quad {\rm and}\quad N g_2 +{g_4\over8\pi}\ln N \ {\rm fixed}\
.$$ 
\section{The $V(\phib \sp{2}) $ in the large $N$ limit: phase transitions}

In higher dimensions something new happens: the possibility of  
phase transitions associated with   
spontaneous 
breaking the $O(N)$ symmetry. In the first part we thus
study the $O(N)$ symmetric $V(\phib^2)$ field theory, in the large $N$
limit to explore the possible phase transitions and identify the corresponding
multicritical points. The action is:
$$S(\phib)= N\int\d^d x \left\{\half \left[ \partial_{\mu} \phib (x)
\right]^{2} +V\left(\phib^2\right) \right\} ,\eqnd{\eactONg}$$
where, as above (Eqs.\eactONgii-\eDPV), 
an implicit cut-off $\Lambda$ is always assumed below. \par
The identity 
\egeniden\ transforms the action into a quadratic form in $\phib$
and therefore the integration over $\phib$ can be  performed. It is convenient
however here to integrate only over 
$N-1$ components, to keep a component of the vector field, which we denote
$\sigma$, in the action. The effective action is then:
$$S_{\rm eff}=N\int\d^d x \left[\half \left(\partial_\mu\sigma\right)^2+
V(\rho)+\half \lambda\left(\sigma^2- \rho\right)\right]+\half
(N-1)\tr\ln(-\Delta+\lambda ).\eqnd{\eactefNg}$$
\subsection The saddle point equations: the $O(N)$ critical point

Let us then write the saddle point equations for a general potential $V$. At
high temperature $\sigma=0$ and $\lambda$ is the
$\phib$-field mass squared. We thus set in general $\lambda=m^2$. With this
notation the three saddle point equations are: 
\eqna\esaddNgen
$$\eqalignno{m^2\sigma&=0\,, & \esaddNgen{a} \cr
V'(\rho)&=\half m^2\,,&\esaddNgen{b}\cr
\sigma^2&=\rho-{1\over (2\pi)^d }\int^{\Lambda} {\d^d k \over k^2
+m^2}\,, & \esaddNgen{c} \cr}$$
with the notation of section 3
$${1\over (2\pi)^d }\int^\Lambda {\d^d k \over k^2
+m^2} \equiv {1\over (2\pi)^d }\int{\d^d k \over D(k^2/\Lambda^2) k^2
+m^2}\equiv B_1(m^2).$$
In the ordered phase $\sigma\ne0$ and thus $m$ vanishes. 
 Equation \esaddNgen{c} has a solution only for $\rho>\rho_c$,
$$\rho_c={1\over (2\pi)^d }\int^\Lambda {\d^d k \over k^2}\,,\ \Rightarrow\
\sigma=\sqrt{\rho-\rho_c}\,.$$
Equation \esaddNgen{b} which reduces to $V'(\rho)=0$ then yields the critical
temperature. Setting
$V(\rho)=U(\rho)+\half r\rho$, we find
$$r_c=-2 U'(\rho_c).$$
To find the magnetization critical exponent $\beta$ we need the relation
between the $r$ and $\rho$ near the critical point. \par
In the disordered phase, $\sigma=0$, equation \esaddNgen{c} relates $\rho$ to
the $\phib$-field mass $m$. It can be rewritten

$$\rho=B_1(m^2)=\Lambda^{d-2}F(m^2/\Lambda^2), \eqnd\enorcritb $$
where
$$F(z)={2\over (4\pi)^{d/ 2} \Gamma({d/ 2}) }\int{k^{d-1}dk
\over k^2D(k^2)+z}\, . \eqnd\eFofz$$

The function $F(z)$ can be written in  an asymptotic expansion
$$F(z)= z^{{d/ 2}-1}\sum_{n=0}^{\infty}
 b_n z^n + \sum_{n=0}^{\infty} c_n z^n.\eqnd\eFofzdef $$
The non-analytic part can be extracted from the representation
$$ F(z)={z^{{d/ 2}-1}\over2\Gamma(d/2) (4\pi)^{d/ 2}}
\int^\infty_0 \d x \int^\infty_0 \d y\, y^{-{d/ 2}}x^{{d/ 2}-1}
\exp\left\{-{x\over 2}D(z{x/ y}) -{y\over 2}\right\}, \eqnd\eFofzINT $$
which gives e.g.~in the case of $D(k^2)=1+d_1 k^2$ the asymptotic expansion
for the non-analytic part:
$$F^{\rm NA}(z)=\Gamma(1-d/2){z^{d/2-1}\over (4\pi)^{d/2}}
\sum_{n=0}^{\infty} {(d_1 z)^n\over n!}
{\Gamma({d\over 2}+2n)\over \Gamma({d\over 2}+n)}\,, \eqnd\eFofzNA$$
and the analytic part is obtained from $F(z)-F^{\rm NA}(z)$.
For $m\ll\Lambda$, $\rho$ approaches $\rho_c$,
$$\rho_c =B_1(0)=\Lambda^{d-2}F(0),$$
and the relation becomes:
$$\rho-\rho_c =-K(d)
m^{d-2}+a(d)m^2\Lambda^{d-4}
+O\left(m^d\Lambda^{-2}\right)
+O\left(m^4\Lambda^{d-6}\right). \eqnd\edmum  $$
For $2<d<4$ (the situation we shall assume below except when stated otherwise)
the $ O\left(m^d\Lambda^{-2}\right)$  from the
non-analytic part dominates the corrections to the leading part of this
expression. 
For $d=4$ instead 
$$\rho-\rho_c=\frac{1}{8\pi^2} m^2\left(\ln m/\Lambda+\ {\rm const.}\right),$$
and for $d>4$ the analytic contribution dominates and 
$$\rho-\rho_c\sim a(d)m^2\Lambda^{d-4}.$$
The constant $K(d)$ is universal:
$$K(d)=-  b_0 =
{1\over(2\pi)^d}\int{\d^d k\over k^2(k^2+1)}=-{\Gamma(1-d/2)
\over(4\pi)^{d/2}}
\,.\eqnd\eKdef$$
The constant $a(d)$ instead depends on the cut-off procedure,
and is given by
$$a(d)= c_1={1\over (2\pi)^d}\int{\d^d k\over k^4}\left(1-{1\over
D^2(k^2)} \right)  .  \eqnd\eadef $$

Let us also define for later purpose the function
$$B_2(p;m^2)={1\over(2\pi)^d}\int{\d^d k\over\left[k^2D(k^2/\Lambda^2)
+m^2\right]
\left[(p-k)^2 D((p-k)^2/\Lambda^2) +m^2\right]},\eqnd\eBmpdef$$
which is up to sign the second derivative with respect
to $\lambda(x)$ of the tr ln term in the effective action. Then for $m$ small
$$\eqalign{B_2(0;m^2)&=-{\d \over\d m^2}B_1(m^2)=-{\d \over\d
m^2}(\rho-\rho_c) \cr
&=\half(d-2)K(d)m^{d-4}-a(d)\Lambda^{d-4}
+O\left(m^{d-2}\Lambda^{-2},m^2\Lambda^{d-6}\right)
.\cr} \eqnd\eBmpad$$
\medskip
{\it Critical point.} In a generic situation $V''(\rho_c)=U''(\rho_c)$ does
not vanish. We thus find in the low temperature phase
$$t=r-r_c\sim -2 U''(\rho_c)(\rho-\rho_c) \ \Rightarrow\ \beta=\half\,.
\eqnd\erminrc$$
This is the case of an ordinary critical point. Stability implies
$V''(\rho_c)>0$ so that $t<0$.\par
At high temperature, in the disordered phase, the $\phib$-field mass $m$ is
given by $2U'(\rho)+ r=m^2$ and thus, using \edmum, at leading order
$$t\sim 2U''(\rho_c)K(d)m^{d-2},$$
in agreement with the result of the normal critical point. Of course the
simplest realization of this situation is to take $V(\rho)$ quadratic, and we
recover the $(\phib^2)^2$ field theory.
\smallskip
{\it The sign of the constant $a(d)$.} A comment concerning the non-universal
constant $a(d)$ defined in
\edmum\ is here in order because, while its absolute value is irrelevant, its
sign plays a role in the discussion of
multicritical points.
Actually this sign is already relevant to
the RG properties of the large $N$ limit of  simple
scalar field theories.
In a $V(\phib^2)={\mu \over 2}\phib^2 +{\lambda \over 4 !}(\phib^2)^2$
theory (in Eq.~\eactONg) it is easy to verify that $a(d)$ is related to
the second coefficient of the large $N$ RG $\beta$-function. If we
call $N\lambda=g\Lambda^{4-d}$ the bare
coupling constant, we indeed find \rbook:
$$\beta(g)=-(4-d)g+\frac{1}{6}(4-d)Na(d)g^2+O\left(1/N\right).\eqnd\betafn$$
It is generally assumed that $a(d)>0$. Indeed, this is what is found near four
dimensions in all regularizations. Then there exists an IR fixed point,
non-trivial zero of the $\beta$-function. 
For the simplest Pauli--Villars's type regularization we have $D(z)>1$ and thus
$a(d)$ is finite and positive in dimensions $2 < d < 4$, but this is not a
universal feature.
Even in case of simple lattice regularizations it has been shown
\ref\rKN{D.A. Kessler and H. Neuberger, {\it Phys. Lett.} 157B (1985) 416.}
that in $d=3$  the sign is arbitrary. We illustrate the ambiguity in the
sign of $a(d)$ in $2 < d < 4$  by explicit examples in appendix A2. 
However, if $a(d)$ is negative
the large $N$ RG has a problem, since the coupling flows in the IR limit
to large values where the large $N$ expansion is no longer reliable. It is
not known whether this signals a real physical problem, or is just an
artifact of the large $N$ limit. \par
Another way of stating the problem is to examine directly the relation
between bare and renormalized coupling constant. Calling $g_{\rm r}
m^{4-d}$ the renormalized 4-point function at zero momentum, we find
$$m^{4-d}g_{\rm r}={\Lambda^{4-d}g\over 1+\Lambda^{4-d}g N B_2(0;m^2)/6}  .
\eqnd\egrenor$$
In the limit $m\ll\Lambda$ the relation can be written
$${1\over g_{\rm r}}={(d-2) N K(d)\over 12}+\left(m\over\Lambda\right)^{4-d}
\left({1\over g}-{N a(d)\over6}\right).\eqnd\egrenor $$
We see that when $a(d)<0$ the renormalized IR fixed point value cannot be
reached by varying $g>0$ for any finite value of $m/\Lambda$.

\subsection Multicritical points

A new situation arises if we can adjust a parameter of the
potential in such a way that $U''(\rho_c)=0$. This can be achieved only if the
potential $V$ is at least cubic. We then expect a tricritical
behavior. Higher critical points can be obtained when more derivatives vanish.
We shall examine the general case though, from the point of view of real phase
transitions, higher order critical points are not interesting because $d>2$
for continuous symmetries and mean-field behavior is then obtained for
$d\ge 3$. The analysis will however be useful in the study of  double scaling
limit.\par
Assuming that the first non-vanishing derivative is $U^{(n)}(\rho_c)$,
we expand further equation \esaddNgen{b}. In the ordered low temperature
phase we now find
$$t=-{2\over (n-1)!}U^{(n)}(\rho_c)(\rho-\rho_c)^{n-1},\ \Rightarrow\
\sigma\propto (-t)^\beta,\quad \beta={1\over2(n-1)}\,  ,  \eqnd\ecritbeta $$
which leads to the exponent $\beta$ expected in the mean field approximation
for such a multicritical point.
We have in addition the condition $U^{(n)}(\rho_c)>0$.\par
In the high temperature phase instead
$$m^2=t+ (-1)^{n-1}{2\over (n-1)!}U^{(n)}(\rho_c)K^{n-1}(d)m^{(n-1)(d-2)}.
\eqnd\emsq$$
For $d>2n/(n-1)$ we find a simple mean field behavior, as expected since we
are above the upper-critical dimension . \par
For $d<2n/(n-1)$ we find a peculiar phenomenon, the term in the r.h.s.\ is
always dominant, but depending on the parity of $n$ the equation has solutions
for $t>0$ or $t<0$. For $n$ even, $t$ is positive and we find
$$m\propto t^\nu,\qquad \nu={1\over(n-1)(d-2)},\eqnd\emoft$$
which is a non mean-field behavior below the critical dimension.
However for $n$ odd (this includes the tricritical point) $t$ must be
negative, 
in such a way that we have now two competing solutions at low temperature.
We have to find out which one is stable. We shall verify below that only the
ordered phase is stable, so that the correlation length of the $\phib$-field
in the high temperature phase always remains finite. Although these dimensions
do not correspond to physical situations because $d<3$ the result is
peculiar and inconsistent with the $\varepsilon$-expansion. 
\par
For $d=2n/(n-1)$ we find a mean field behavior without logarithmic
corrections, provided one condition is met:
$${2\over(n-1)!}U^{(n)}(\rho_c)K^{n-1}\left(2n/(n-1)\right)<1\,,\qquad
K(3)=1/(4\pi).\eqnd\etriccond$$
We examine, as an example, in more details the tricritical point below.
We will see that the special point
$${2\over(n-1)!}U^{(n)}(\rho_c)K^{n-1}\left(2n/(n-1)\right)=1\,,
\eqnd\eendtric$$
has several peculiarities \rBMB. 
In what follows we call $\Vc$ this special value of $U^{(n)}(\rho_c)$.
\medskip
{\it Discussion.} In the mean field approximation the function
$U(\rho)\propto\rho^n$ is not bounded from below for $n$ odd, however $\rho=0$
is the minimum because by
definition $\rho\ge0$. Here instead we are in the situation where $U(\rho)
\sim (\rho-\rho_c)^n$ but $\rho_c$ is positive. Thus this extremum of the
potential is likely to be unstable for $n$ odd. To check the global
stability requires further work. The question is whether such multicritical
points can be studied by the large $N$ limit method.
\par
Another point to notice concerns renormalization group: The $n=2$ example is
peculiar in the sense that the large $N$ limit exhibits a non-trivial IR fixed
point. For higher values of $n$ no coupling renormalization arises in the
large $N$ limit and the IR fixed point remains pseudo-gaussian. We are in a
situation quite
similar to usual perturbation theory, the $\beta$ function can only be
calculated perturbatively in $1/N$ and the IR fixed point is outside the
perturbative regime.
\subsection Local stability and the mass matrix

The matrix of the general second partial derivatives of the effective action
is:
$$N\pmatrix{p^2+m^2 & 0 & \sigma \cr
0 & V''(\rho) & -\half \cr
\sigma & -\half & -\frac{1}{2}B_2(p;m^2)  \cr}  ,  \eqnd\ematrix$$
where $B_2(p;m^2)$ is defined in \eBmpdef.\par
We are in position to study the local stability of the critical points.
Since the integration contour for $\lambda=m^2$ should be parallel to the
imaginary axis, a necessary condition for stability is that the determinant
remains negative.
\smallskip
{\it The disordered phase.} Then $\sigma=0$ and thus we have only to study
the $2\times2$ matrix $\bf M$ of the $\rho,m^2$  subspace. Its determinant
must remain negative which implies
$$\det{\bf M}<0\ \Leftrightarrow\
2V''(\rho)B_2(p;m^2)+1>0\,.\eqnd\estable $$
For Pauli--Villars's type regularization the function 
$B_2(p;m^2)$ is decreasing
so that this condition is implied by the condition at zero momentum
$$\det{\bf M}<0\ \Leftarrow\ 2V''(\rho)B_2(0;m^2)+1>0\,.$$
For $m$ small we use Eq.~\eBmpad~and
at leading order the condition becomes:
$$ K(d)(d-2)m^{d-4}V''(\rho)+1>0\,.$$
This condition is satisfied by a normal critical point since $V''(\rho_c)>0$.
For a multicritical point, and taking into account equation \edmum\ we find:
$$(-1)^n {d-2\over (n-2)!}K^{n-1}(d)m^{n(d-2)-d}V^{(n)}(\rho_c)+1>0\,.
\eqnd\multicr $$ 
We obtain a result consistent with our previous analysis: For $n$ even it is
always satisfied. For $n$ odd it is always satisfied above the critical
dimension and never below. At the upper-critical dimension we find a condition
on the value of $V^{(n)}(\rho_c)$ which we recognize to be identical to
condition \etriccond\ because then $2/(n-1)=d-2$.
\smallskip
{\it The ordered phase.} Now $m^2=0$ and the determinant $\Delta$ of the
complete matrix is:
$$-\Delta>0\ \Leftrightarrow\
2V''(\rho)B_2(0;p^2)p^2+p^2+4V''(\rho)\sigma^2>0\,.\eqnd\eorder$$
We recognize a sum of positive quantities, and the condition is always
satisfied. Therefore in the case where there is a competition with a
disordered saddle point only the ordered one can be stable.
\section The scalar bound state

In this section we study the limit of stability in the disordered phase
($\sigma=0$). This is a problem which only arises when $n$ is odd, the first
case being provided by the tricritical point.
The mass-matrix has then a zero eigenvalue which corresponds to the appearance
of a new massless excitation other than $\phib$. Let us denote by $\bf M$ the
$\rho,m^2$ $2\times2$ submatrix. Then
$$\det{\bf M}=0\ \Leftrightarrow\ 2V''(\rho)B_2(0;m^2)+1=0\,.$$
In the two-space the corresponding eigenvector has components 
$(\half,V''(\rho))$.
\subsection The small mass $m$ region

In the small $m$ limit the equation can be rewritten in terms of the
constant $K(d)$ defined in \eKdef:
$$ K(d)(d-2)m^{d-4}V''(\rho)+1=0\,.\eqnd\ephitwoa $$
Equation \ephitwoa\ tells us that $V''(\rho)$ must be small. We are thus
close to a multicritical point. Using the result of the stability analysis
we obtain
$$(-1)^{n-1}{d-2\over (n-2)!}K^{n-1}(d)m^{n(d-2)-d}V^{(n)}(\rho_c)=1\,.
\eqnd\estabil$$
We immediately notice that this equation has solutions only for $n(d-2)=d$,
i.e.\ at the critical dimension. The compatibility then fixes the value of
$V^{(n)}(\rho_c)$. We again find the point \eendtric, $V^{(n)}(\rho_c)=\Vc$.
If we take into account the leading correction to the small $m$ behavior
we  find  instead:
$$V^{(n)}(\rho_c)\Vc^{-1}-1\sim(2n-3){a(d)\over K(d)}\left({m\over\Lambda}
\right)^{4-d}.\eqnd\estabil  $$
This means that when $a(d)>0$ there exists a small region
$V^{(n)}(\rho_c)>\Vc$ where the vector
field is massive with a small mass $m$ and the bound-state massless. The
value $\Vc$ is a fixed point value.
\medskip
{\it The scalar field at small mass.} We want to extend the analysis to a
situation where the scalar field has a small but non-vanishing mass $M$ and
$m$ is still small. The goal is in particular to explore the neighborhood of
the special point \eendtric.
Then the vanishing of the determinant of $\bf M$ implies
$$1+2V''(\rho)B_2(iM;m^2)=0\,.\eqnd\emassless$$
Because  $M$ and $m$ are small, this equation still implies
that $\rho$ is close to a point $\rho_c$ where $V''(\rho)$ vanishes.
Since reality imposes $M<2m$, it is easy to verify that this equation
has also solutions for only the critical dimension. Then
$$V^{(n)}(\rho_c)f(m/M)=\Vc\,,\eqnd\eOmegaC$$
where we have set:
$$f(z)=\int_0^1\d x\left[1+(x^2-1)/(4z^2)\right]^{d/2-2},\qquad \half<z\,.
\eqnd\efofz$$
In three dimensions it reduces to:
$$f(z)=z\ln\left({2z+1\over 2z-1}\right)\,.$$
$f(z)$ is a decreasing function which diverges for $z=\half$ because $d\le3$.
Thus we find solutions in the whole region $0<V^{(n)}(\rho_c)<\Omega_c$,
i.e.\ when the multicritical point is locally stable.\par
Let us calculate the propagator near the pole. We find the matrix $\bf \Delta$
$${\bf \Delta}={2\over G^2}\left[N\left.{\d B_2(p;m^2)\over \d
p^2}\right \vert_{p^2=-M^2}\right]^{-1}{1\over p^2+M^2}\pmatrix{1 &G
\cr G  &  G^2 \cr},\eqnd\edeltapole$$
where we have set
$$ G={2(-K)^{n-2}W^{(n)} \over  (n-2)! } m^{4-d}\,. $$
For $m/M$ fixed the residue goes to zero with $m$ as 
$m^{d-2}$ because the derivative of $B$ is of the order of $m^{d-6}$. 
Thus the bound-state decouples on the multicritical line.
\subsection The scalar massless excitation: general situation

Up to now we have
explored only the case where both the scalar field and the vector field
propagate. Let us now relax the latter condition, and examine what happens
when $m$ is no longer small. The condition $M=0$ then reads
$$2V''(\rho_s)B_2(0;m^2)+1=0\,   $$
together with
$$m^2=2V'(\rho_s),\qquad \rho_s= B_1(m^2)\,. \eqnd\GapEq $$
An obvious remark is: there exist solutions only for $V''(\rho_s)<0$, and
therefore the ordinary critical line can never be approached. In terms
of the function $F(z)$ defined by equation \enorcritb\ the equations
can be rewritten
$$\rho_s=\Lambda^{d-2} F(z),\quad z=2V'(\rho_s)\Lambda^{-2},\quad
2\Lambda^{d-4} V''(\rho_s)F'(z)=1\,.$$
The function $F(z)$ in Pauli--Villars's regularization is a decreasing
function.
In the same way $-F'(z)$ is a positive decreasing function.\par
The third equation is the condition for the two curves corresponding to the
two first ones become tangent. For any value of $z$ we can find potentials and
thus solutions. Let us call $z_s$ such a value and specialize to cubic
potentials. Then
$$\rho_s=\Lambda^{d-2}F(z_s)~~,$$ 
$$\quad V(\rho)=V'(\rho_s)(\rho-\rho_s)+\half
V''(\rho_s)(\rho-\rho_s)^2+
\frac{1}{3!}V^{(3)}(\rho_s)(\rho-\rho_s)^3,\eqnd\Vofrho $$
which yields a two parameter family of solutions. For $z$ small we see that
for $d<4$ the potential becomes proportional to $(\rho-\rho_c)^3$.

\section Stability and double scaling limit

In order to discuss in more details the stability issue and the double scaling
limit we 
now construct the effective action for the scalar bound state. We consider
first only the massless case. We only need the action in the IR limit, and in
this limit we can integrate out the vector field and the second massive
eigenmode. 
\medskip
{\it Integration over the massive modes.} 
As we have already explained in section 3 we can integrate over one of
the fields, the second being fixed, and we need only the result at leading
order. Therefore we replace in the functional integral
$$\e^Z=\int [\d\rho\d\lambda]\exp\left[-\frac{N}{2}\tr \ln(-\del^2 + \lambda)
+N\int \d^dx\left(- V(\rho) +\half \rho \lambda\right)\right],\eqnd\eZ$$
one of the fields by the solution of the field
equation. It is useful to discuss the effective potential of the
massless mode first. 
This requires calculating the action only for constant fields
it is then simpler to eliminate $\lambda$. We 
assume in this section that $m$ is small (the vector propagates). 
For $\lambda\ll\Lambda$ the $\lambda$-equation reads ($d<4$)
$$\rho-\rho_c=-K(d)\lambda^{(d-2)/2}.  \eqnd\erholambda $$
It follows that the resulting potential $W(\rho)$, obtained from Eq.~\eWeffrho
~is 
$$W(\rho)=V(\rho)+{d-2\over 2d (K(d))^{2/(d-2)}}(\rho_c-\rho)^{d/(d-2)}.
\eqnd\effUrho$$
In the sense of the double scaling limit the criticality condition is
$$W(\rho)=O\bigl((\rho-\rho_s)^n\bigr).$$
It follows
$$V^{(k)}(\rho_s)=- \half K^{1-k}(d){\Gamma\bigl(k-d/(d-2)\bigr)\over
\Gamma\bigl(-2/(d-2)\bigr)}m^{d-k(d-2)}\quad 1\le k\le n-1\, .$$
For the potential $V$ of minimal degree we find
$$W(\rho)\sim \frac{1}{2 n!} K^{1-n}(d){\Gamma\bigl(n-d/(d-2)\bigr)\over
\Gamma\bigl(-2/(d-2)\bigr)}m^{d-n(d-2)}(\rho-\rho_s)^n .$$
\medskip
{\it The double scaling limit.}
We recall here that quite generally one verifies that a non-trivial double
scaling limit may exist only if the resulting field theory of the massless
mode is 
super-renormalizable, i.e.~below its upper-critical dimension $d=2n/(n-2)$,
because perturbation theory has to be IR divergent. Equivalently, to
eliminate $N$ from the critical theory, one has to rescale
$$\rho-\rho_s\propto N^{-2\theta}\varphi\,,\quad x\mapsto x N^{(n-2)\theta}
\quad {\rm with}\ 1/\theta=2n-d(n-2),$$ 
where $\theta$ has to be positive.
\par
We now specialize to dimension three, since $d<3$ has already been
examined, and the expressions above are valid  only for $d<4$.
The normal critical point ($n=3$), which leads to a $\varphi^3$ field theory, 
and can be obtained for a quadratic potential $V(\rho)$  (the $(\phib^2)^2$)
has been discussed elsewhere \rVecd. We thus concentrate on the next critical
point $n=4$ where the minimal potential has degree three.
\medskip
{\it The $d=3$ tricritical point.}
The potential $W(\rho)$ then becomes
$$W(\rho)=V(\rho)+\frac{8\pi^2}{3}(\rho_c-\rho)^3. \eqnd\effWthreeD$$
If the potential $V(\rho)$ has degree larger than three, we obtain
after a local expansion and a rescaling of fields,
$$\rho-\rho_s= ({-1\over 32\pi^2 \rho_c})
(\lambda-m^2)\propto \varphi/N\,,\quad x\mapsto Nx\,,\eqnd\rescal$$
a simple super-renormalizable $\varphi^4(x)$ field theory. 
 If we insist
instead that the initial theory should be renormalizable, then we remain with
only one candidate, the renormalizable $(\phib^2)^3$ field theory, also
relevant for the tricritical phase transition with $O(N)$ symmetry breaking.
Inspection of the potential $W(\rho)$ immediately shows a remarkable feature:
Because the term added to $V(\rho)$ is itself a polynomial of degree three,
the critical conditions lead to a potential $W(\varphi)$ which vanishes
identically. This result reflects the property that 
the two saddle point equations  ($\del S/\del \rho = 0 ~,~  
\del S/\del \lambda = 0$ in   Eqs.~\esaddNgen) are proportional 
and thus have a continuous one-parameter family of solutions. This results in
a flat effective potential for $\varphi(x)$.
The effective action for $\varphi$ depends only on the derivatives
of $\varphi$, like in the $O(2)$ non-linear $\sigma$ model.
We conclude that no non-trivial double scaling limit can 
be obtained\footnote*{
It was pointed out that another sort of obstacle is present for
$(\phib^2)^2$ model in $d=2$
\ref\DiVMo{P. Di Vecchia and M. Moshe, {\it Phys.~Lett.~}B300 (1993) 49.}\ 
and $d=4$ \ref\Sch{H.~J.~Schnitzer, {\it Mod.~Phys.~Lett.~}A7 (1992) 2449.}.}
under these conditions.
In three dimensions with a $(\phib^2)^3$ interaction we can generate at
most a normal critical point $n=3$, but then a simple $(\phib^2)^2$
field theory suffices.

The ambiguity of the sign of $a(d)$ discussed in section 4 and in  Appendix A2
has an interesting appearance in $d= 3$ in the small $m^2$ region.  
If one keeps the
extra term proportional to $a(d)$ in Eq.~\effUrho~we have
$$W(\rho)=V(\rho)+{8\pi^2 \over 3}(\rho_c-\rho)^3
+{a(3)\over\Lambda} 4\pi^2 (\rho_c-\rho)^4.$$
Using now Eq.~\erholambda~and, as mentioned in section 5, 
the fact that in the small $m^2$ region the potential is proportional to
$(\rho-\rho_c)^3$ we can solve for $m^2$.
Since $m^2>0$ the appearance of a phase with small mass depends
on the sign of $a(d)$. Clearly this shows a
non-commutativity of the limits of $m^2/\Lambda^2 \to 0 $ and $ N \to \infty$.
The small  $m^2$ phase can be reached by a special tuning  and
cannot be reached with an improper sign of $a(d)$.
Calculated in this way, $m^2$ can be made proportional to the deviation  of
the  coefficient of $\rho^3$ in $V(\rho)$ from its critical  value $16\pi^2$. 
\section Conclusions 

This is a study of several subtleties in the phase structure of  $O(N)$ vector
models around multicritical points of odd and even orders.
One of the main topics is the understanding of the multicritical behavior
of these models  at their critical dimensions and the effective field theory
of the $O(N)$-singlet bound state obtained in the $N \to \infty$, ~$g \to g_c$
correlated limit. It is pointed out (in contrast to previous studies) 
that the integration over massive $O(N)$ singlet modes is essential 
in order to extract the correct effective field theory of
the small mass scalar excitation. 
After performing this integration, it has been established here that
the double scaling limit of $(\phib^2)^n$ vector model 
in its critical dimension $d=2n/(n-1)$ can 
result in a theory of a free massless $O(N)$ singlet bound state.  
This fact is a consequence of the existence of 
flat directions at the scale invariant multicritical point 
in the effective action. In contrast to the case $d < 2n/(n-1)$
where IR singularities compensate powers of $1/N$ in the double scaling 
limit,  at  $d=2n/(n-1)$ there is no such compensation and only a 
noninteracting effective field theory  of the massless bound state is left.
\par
Another interesting issue in this study is the ambiguity of the
sign of $a(d)$.
The coefficient of $m^2\Lambda^{d-4}$ denoted by $a(d)$ in the expansion
of the gap equation in Eqs.~\esaddNgen{c} and \edmum~seems to have a
surprisingly important role in the approach to the continuum limit ($\Lambda^2
\gg m^2$). The existence of an IR fixed point  at $g \sim O(N^{-1}),$ as seen
in the $\beta$ function for the unrenormalized coupling constant in
Eq.~\betafn, depends on the sign of $a(d)$. Moreover, as seen in section 3.1 
the existence of a phase with a small mass $m$ for the $O(N)$ vector quanta
and a massless $O(N)$ scalar depends also on the sign of
$a(d)$. It may very well be that the importance of 
the sign of $a(d)$ is a mere 
reflection of the limited  coupling constant space used to described the
model. 
This is left here as an open question that deserves a detailed renormalization
group or lattice simulation study in the future.
\vfill\eject
\medskip
\noindent {\bf Acknowledgments}
\medskip
This  research  has been supported in part (MM) by the BSF,  the 
 VPR Fund for promotion of research
at the Technion,  and the GIF, (SN) 
by the Israeli Council for Higher Education
and by JSPS Postdoctoral Fellowships for Research Abroad.

\vskip 1.5cm

\def\appendixname{}
\appendix{}

\vskip -2cm

\section Double scaling limit: $d=0$

We consider the sum $Z$ of connected Feynman diagrams in $d=0$ dimension:
$$\e^Z = \int \d^N \phib \e^{-NV(\phib^2)} = {\pi^{N/2}\over
\Gamma(N/2)}\int^{\infty}_{0}{\d x\over x} \e^{-N[V(x)- 
\ln (x)/2]},\eqnd\eZN $$ 
where $x$ is normalized by $V(x)=\half x+O(x^2)$.\par
We define
$$W(x)=V(x)-\half\ln(x) .$$
The saddle point equation reads
$$W'(x_s)=0\,.$$
A critical point of order $n$ is defined by the conditions:
$$ W(x) - W(x_s)
\mathop{\sim}_{x\to x_s} 
 W^{(n)}(x_s){\left(x-x_s\right)^n\over
n!}$$
The critical potential $V_n(x)$ of lowest degree thus satisfies
$$ V'_{n}(x)=\left[1-(1-x/(n-1))^{n-1}\right]/(2x),$$
and therefore
$$V_{n}(x)=\half\int_0^{x/(n-1)}{\d y\over y}\left[1-(1-y)^{n-1}\right]. $$
and
$$ W'_{n}
\mathop{\sim}_{x\to n}-\frac{1}{2(n-1)}(1-x/(n-1))^{n-1} .$$
If we now change the variable, setting 
$$(1-x/(n-1))N^{1/n}=z\,, \eqnd\evarxnz $$ 
we find
$$\e^{Z_c} \sim$$
$$ {\pi^{N/2}N^{-1/n}\over \Gamma(N/2)}\exp[\half N(\ln
(n-1)-1-\half-\cdots -\frac{1}{n-1})] \int\d
z\,\e^{-z^{n}/2n + O(N^{-1/n}z^{n+1})}. $$
Adding to the critical potential all relevant perturbations, and shifting
the saddle point back to $x=n-1$, we obtain
$$V(x)=V_{n}(x)+\sum_{k=1}^{n-1}{v_k\over 2k}\left[ (1-x/(n-1))^k-
\frac{k}{n}(1-x/(n-1))^{n}\right] . \eqnd\escalpot $$
After the change of variable \evarxnz, we find at leading order for $N$ large
$$N\bigl(W(x)-W(n)\bigr)=\frac{1}{2n}z^{n}+
\sum_{k=1}^{n-1} {v_k\over 2k}\left[ N^{1-k/n}z^k-\frac{k}{n}z^{n}\right]
.$$ 
We see that a double scaling limit is reached only when the coefficients $v_k$
go to zero for large $N$ with $v_k N^{1-k/n}=u_k $ fixed. This implies
that the coupling constants, i.e.~the coefficients of $V(x)$ in the expansion
in powers of $x$ must approach the critical values corresponding to the
potential $V_{n}$ with a well-defined behaviour as a function of $N$.
Note also that since for all $v_k=0$ several solutions of the saddle point
equation coalesce, the critical potential corresponds to a singularity
of the large $N$ partition function in the space of coupling constants.
Actually if we expand $Z$ in powers of $1/N$, 
$$Z=\sum_{h=1}N^{1-h} Z_h \,,$$ 
all terms $Z_h$ are singular at this point. This can be most easily seen
by keeping only the most relevant term proportional to $v_1$. Then one
finds
$$Z_h\propto v_1^{(1-h)n/(n-1)} .$$
The scaling behaviour of $v_1$ as a function of $N$ uses this singular
behaviour to compensate the powers of $N$. Finally by tuning the coupling
constants to reach a singularity we approach the radius of convergence
of the perturbative expansion at $h$ fixed, and thus enhance high order
Feynman diagrams. Thus one reaches
a dense configuration of contributing Feynman graphs which is the analog
of the sum over surfaces in $O(N)$ matrix models. In the
$O(N)$ vector model this is an expansion over `randomly branched polymers'.
The scaling partition function is obtained by simplifying
the integrand
$$N\bigl(W(x)-W(n-1)\bigr)\sim \frac{1}{2n}z^{n}+
\sum_{k=1}^{n-2} {u_k\over 2k} z^k ,$$
because the $v_k$ goes to zero and the term $k=n-1$ can be eliminated by
shifting $z$.  Finally, if we keep only the most relevant perturbation term
proportional to $v_1$ the double scaled partition function is
given by a generalized Airy function
$$ Z(u_1)=\int \d  z \exp \left( 
-\half\bigl( {\frac{z^{n}}{n}} + u_1 z\bigr) \right).   $$

\section Regularization and sign of the $\beta$-function

Here we give a few explicit examples which show 
the regularization dependence of the
non-universal constant $a(d)$ defined in \edmum.
\medskip
{\it Pauli--Villars regularization.}
For the regularization \eDPV\ it is given by
$$a(d)={1\over (2\pi)^d}\int{\d^d k\over k^4}\left(1-{1\over
D^2(k^2)} \right).$$
Let us consider for example
$$D(k^{2})=1+\alpha k^{2}+\beta k^{4}. $$
$D(k^2)>0$ implies $4\beta>\alpha^2$.\par
If $\alpha >0$ then clearly for every $k$, $D(k^{2})>1$, and  $a(d)$
is positive.
However if $\alpha<0$ it is possible to choose the parameters in such a way
that $a(d)$ will change sign; especially if one takes $\alpha$ close to
$-2\sqrt{\beta}$. Possible values of $a(d)$ are exhibited in Fig.~1 below.

\vskip -2cm
\epsfxsize=8cm
\centerline{\epsfbox{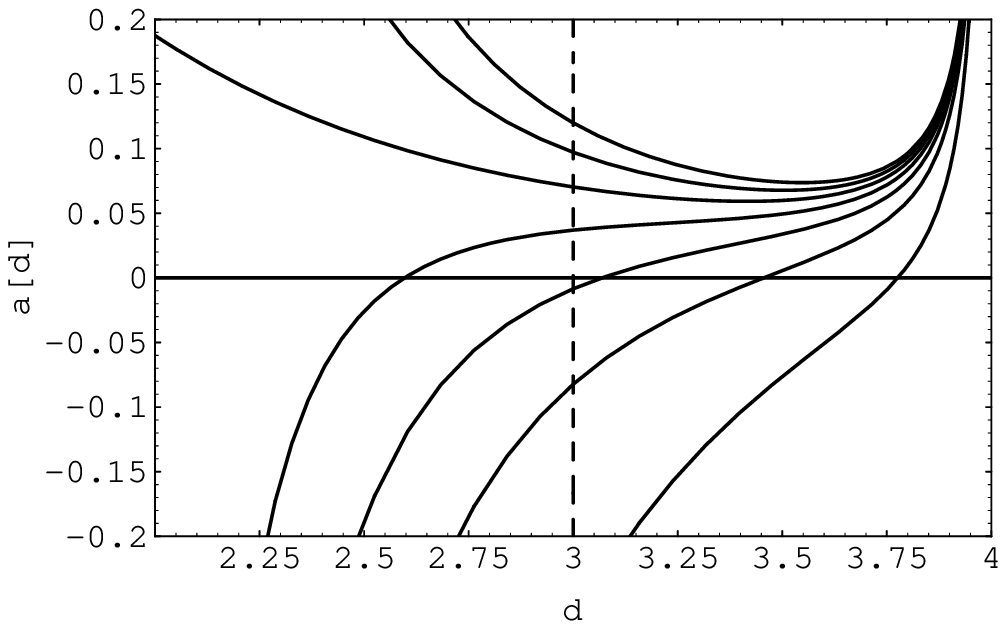}}
\vskip -1cm
\centerline{Fig.~1:\  Values of $a(d)$ in Pauli--Villars regularization.}
\centerline{
(From top to bottom: $\alpha=0.8,\ 0.4,\ 0,\ -0.4,\ -0.8,\ -1.2,\ -1.6$ and
$\beta=1$.) }

\bigskip
{\it Lattice regularization.}
We extend here the calculations of \rKN\ to more general lattice
regularizations, to understand how general this sign property is.
\par
An alternative representation to Eq.~\enorcritb\ is:
$$\eqalign{\rho&=\half \int_0^{\infty}\d\alpha\e^{-
\alpha m^{2}/2}f(\alpha)\cr
f(\alpha)&=\int_{-\pi}^{\pi}\frac{\d^{d}k}{(2\pi)^{d}}\e^{-\alpha
Q(k)/2},\cr}$$ 
where $(Q(k))^{-1}$ is the massless free field propagator.
Using the asymptotics of $f(\alpha )$
$$f(\alpha)\mathop{\sim}_{\alpha\rightarrow\infty}(2\pi\alpha)^{-\frac{d}{2}}
\left[1+O(\frac{1}{\alpha})\right],$$
then
$$\rho-\rho_{c}=-K(d)m^{d-2}+\half \int_{0}^{\infty}\d\alpha
\left(\e^{-\alpha m^{2}/2}-1\right)\left(f(\alpha)-
(2\pi\alpha)^{-d/2}\right) . $$
Therefore $a(d)$ in Eq.\edmum\ reads:
$$a(d)=-\frac{1}{4}\int_{0}^{\infty}\d\alpha\,\alpha
\left(f(\alpha)-(2\pi\alpha)^{-d/2}\right).$$
$a(d)$ depends on the regularization used for the lattice action
$$S(\phib )=S_{K}(\phib )+N\sum_{x}\;[V(\phib ^{2})]$$
where
$$
S_{K}(\phib)=\half N(2\pi)^d\int^{\pi}_{-\pi}\d^{d}k\,
{\tilde\phib}(-k)Q(k)
{\tilde\phib}(k),\ \ 
\phib(x) \equiv \int^{\pi}_{-\pi}
\d^{d}k\,\e^{ikx}{\tilde\phib}(k) .$$
The lattice spacing is taken to be 1.

If only nearest-neighbor interactions are included:
$$Q(k)=4\sum_{\mu=1}^{d}\sin^{2}(\frac{k_{\mu}}{2}),$$
then
$$f(\alpha)=(\e^{-\alpha}I_{0}(\alpha))^d$$
where  $ I_{0}(\alpha)$ is a modified Bessel function.

In the following regularization which includes
contributions from next to nearest neighbors and
next to next to nearest neighbors the dimension $d$ can
be varied continuously:
$$\eqalign{S_{K}/N& =  -\frac{1}{2}\sum_{x}\phi(x)\left[(1-r)\sum_{\mu=1}^{d}
\bigl(\phi(x+\hat{\mu})+\phi(x-\hat{\mu})-2\phi(x)\bigr)\right. \cr
&\quad \left.+\frac{1}{4}r\sum_{\mu=1}^{d}
\bigl(\phi(x+2\hat{\mu})+\phi(x-2\hat{\mu})-2\phi(x)\bigr)\right].\cr} $$
The inverse of the free field propagator is now:
$$Q(k;r)=\sum_{\mu =1}^{d}\left(4 \sin^{2}\frac{k_{\mu}}{2}
-4r\sin^{4}\frac{k_{\mu}}{2}\right).$$
Values of $a(d)$ in this regularization are exhibited in Fig.~2 below.

\vskip -2cm
\epsfxsize=8cm
\centerline{\epsfbox{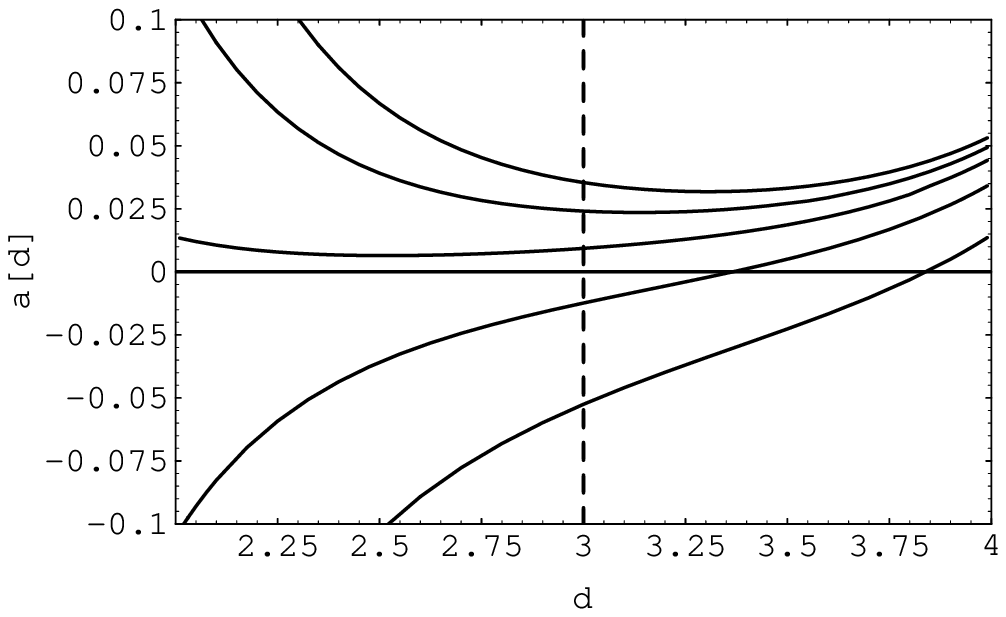}}
\vskip -1cm
\centerline{Fig.~2:\  Values of $a(d)$ in lattice regularization.}
\centerline{
(From top to bottom: $r=-1.2,\ -0.8,\ -0.4,\ 0,\ 0.4$.) }

At $d=3$, 
including all next-to-nearest-neighbors 
we have:
$$\eqalign{S_{K}&/N=-\frac{1}{2}\sum_{x}\phi(x)\left[(1-r)\sum_{\mu=1}^{3}
\bigl(\phi(x+\hat{\mu})+\phi(x-\hat{\mu})-2\phi(x)\bigr)\right. \cr
&\quad\left.+\!\frac{1}{4}r\!\sum_{3\geq\mu >\nu\geq 1}\!\!\bigl(\phi(\!x\!
+\!\hat{\mu}\!+\!
\hat{\nu}\!)\!+\!\phi(\!x\!+\!\hat{\mu}\!-\!\hat{\nu}\!)\!+\!\phi(\!x\!-\!
\hat{\mu}\!+\!\hat{\nu}\!)\!+\!\phi(\!x\!-\!\hat{\mu}\!-\!\hat{\nu}\!)\!-\!4
\phi(\!x\!)\bigr)\right]\cr} $$
and thus
$$Q_{2}(k;r)=4\sum_{\mu=1}^{3}\sin^{2}(\frac{k_{\mu}}{2})-2r\left[\left
(\sum_{\mu=1}^{3}
\sin^{2}(\frac{k_{\mu}}{2})\right)^{2}-\sum_{\mu=1}^{3}
\sin^{4}(\frac{k_{\mu}}{2})\right].$$
Once again, one finds that as the parameter $r$ is growing, $a(d)$ is 
changing its sign from $a(d) > 0$
to $a(d) < 0$ ( we have $a(d) =0.002~,~-0.0001~,~-0.07 $ for $r= -0.6~,~
-0.5~,~ 0.9$ respectively ).

\vfill\eject
\listrefs

\end